\documentclass[prl,aps,twocolumn,showpacs]{revtex4}
\usepackage[dvips]{graphicx}
\usepackage{dcolumn}%
\usepackage{amsmath}%
\usepackage{amsfonts}%
\usepackage{amssymb}
\providecommand{\U}[1]{\protect\rule{.1in}{.1in}}
\newcommand{\be}{\begin{equation}}
\newcommand{\ee}{\end{equation}}
\newcommand{\bea}{\begin{eqnarray}}
\newcommand{\eea}{\end{eqnarray}}
\newcommand{\nn}{ \nonumber}
\newcommand{\ds}{\displaystyle}
\begin{document}

\title{Inelastic electron transport through molecular junctions}

\author{ Natalya A. Zimbovskaya}%

\affiliation{Department of Physics and  Electronics, University of Puerto Rico-Humacao, \\ CUH Station, Humacao, PR 00791, USA}
\affiliation{Institute for Functional Nanomaterials, \\ University of Puerto Rico, San Juan, PR 00931, USA}  

\pacs{71.18.+y, 71.20-b, 72.55+s}

\begin{abstract}
 Currently, molecular tunnel junctions are recognized as important active elements of various nanodevices. This gives a strong motivation to study  physical mechanisms controlling electron transport through molecules. Electron motion through a molecular bridge is always somewhat affected by the environment, and the interactions with the invironment could change the energy of the traveling electron. Under certain conditions these inelastic effects may significantly  modify electron transport characteristics. In the present work we describe inelastic and dissipative effects in the electron transport occurring due to the molecular bridge vibrations and stochastic thermally activated ion motions. We intentionally use simple models and computational techniques to keep a reader focused on the physics of inelastic electron transport in molecular tunnel junctions. We consider electron-vibron interactions and their manifestations in the inelastic tunneling spectra, polaronic effects and dissipative  electron transport. Also, we briefly discuss long-range electron transfer reactions in macromolecules and their relation to the electron transport through molecular junctions.
\vspace{6mm}

\bf {\it Keywords:}  molecular tunnel junctions, electron-phonon interactions, inelastic electron tunneling spectrum, polaronic conduction, electron transfer reactions.


\end{abstract}
\date{\today}
\maketitle

\begin{tabular}{c l r}
& \hspace{20mm} \large \bf Contents &  
\\  \\
\bf  I. &\bf Introduction  &  1
\\  \\
\bf II. & \bf  Coherent transport & 2
 \\ \\
\bf  III. & \bf  Buttiker model for inelastic transport   & 5
\\ \\
\bf  IV. & \bf  Vibration-induced inelastic effects & 6
\\ \\
\bf V. & \bf Dissipative transport  & 9
\\ \\
\bf VI. & \bf Polaron effects: Hysteresis, switching & \\
          &  \bf   and negative differential resistance   & 11
\\ \\
\bf VII. &\bf Molecular junction conductance and   &\\
          &  \bf  long range electron transfer reactions   & 12
\\  \\
\bf VIII. &\bf Concluding remarks   & 15
\\ \\
    &  \bf References   & 16 \\
\end{tabular}

\subsection{I. Introduction}

Molecular electronics is known to be one of the most promising developments in nanoelectronics, and the past decade has seen an extraordinary progress in this field (Aviram 2002, Cuniberti 2005, Nitzan 2001). Present activities on molecular electronics reflect the convergence of two trends in the fabrication of nanodevices, namely, the top-down device miniaturization through the lithographic methods and bottom-up device manufacturing through atom-engineering and self-assembly approach. The key element and basic building block of molecular electronics is a junction including two electrodes (leads) linked by a molecule, as schematically shown in the Fig. 1. Usually, the electrodes are microscopic large but macroscopic small contacts which could be connected to a battery to provide the bias voltage across the junction. Such a junction may be treated as a quantum dot coupled to the charge reservoirs. The discrete character of  energy levels on the dot (molecule) is combined with nearly continuous energy spectra on the reservoirs (leads) occurring due to their comparatively large size.

\begin{figure}[t] 
\begin{center}
\includegraphics[width=7.5cm,height=9cm,angle=-90]{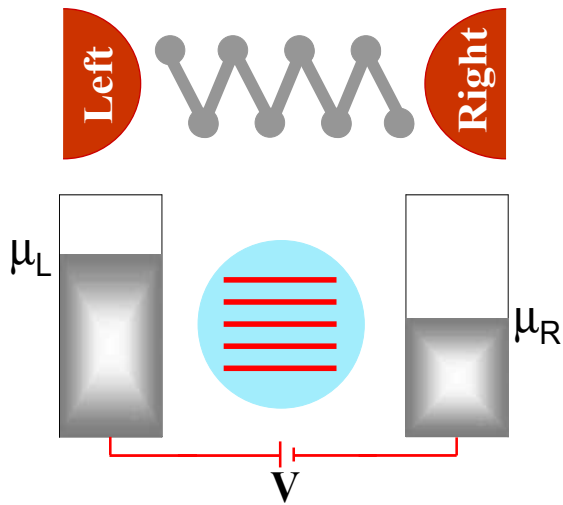}
\caption{(Color online) Top panel: Schematic drawing of a junction including two electrodes and a molecule in between. Bottom panel: When the voltage is applied across the junction electrochemical potentials $\mu_L$ and $\mu_R$ differ, and the conduction window opens up.
}
 \label{rateI}
\end{center}\end{figure}

When the voltage is applied, an electric current flows through the junction. Successful transport experiments with molecular junctions (Ho 2002, Lortscher 2007, Park 2000, Poot 2006, Reicher 2002, Smit 2002, Yu 2004) confirm their significance as active elements in nanodevices. These include applications as rectifiers (molecular diodes), field effect transistors (molecular triods), switches, memory elements and sensors. Also, these experiments emphasize the importance of thorough analysis of the physics underlying
electron transport through molecular junctions. Detailed understanding of the electron transport at the molecular scale is a key step to future device operations. Theory of electron transport in molecular junctions is being developed in the last two decades, and main transport mechanisms are currently elucidated in general terms (Datta 2005, Imry 1999).
However, progress of experimental capabilities in the field of molecular electronics brings new theoretical challenges causing further development of the theory.

Speaking of transport mechanisms, it is useful to make a distinction between the elastic electron transport when the electron energy remains the same as it travels through the junction, and inelastic transport processes when the electron undergoes energy changes due to its interactions with the environment. There are several kinds of processes bringing inelasticity in the electron transport in mesoscopic systems including molecular junctions. Chief among these are electron-electron and electron-phonon scattering processes. These processes may bring significant inelastic effects modifying transport properties of molecular devices and charging, desorption and chemical reactions as well. To keep this Chapter at a reasonable length we concentrate on the inelastic effects originating from electron-phonon interactions.

In practical molecular junctions the electron transport is always accompanied by nuclear motions in the environment. Therefore the conduction process is influenced by the coupling between electronic and vibrational degrees of freedom. Nuclear motions underlie the interplay between the coherent electron tunneling through the junction and inelastic thermally assisted hopping transport (Nitzan 2001). Also, electron-phonon interactions may result in polaronic conduction (Galperin 2005, Gutierrez 2005, Kubatkin 2003, Ryndyk 2008), and they are directly related to the junction heating (Segal 2003) and to some specific effects such as alterations in both shape of the molecule and its position with respect to the leads (Komeda 2002, Mitra 2004, Stipe 1999). The effects of electron-phonon interactions may be manifested in the inelastic tunneling spectrum which presents the second derivative of the current in the junction $d^2I/dV$ versus the applied voltage $V.$ The inelastic electron tunneling microscopy has proven to be a valuable method for identification of molecular species within the conduction region, especially when employed in combination with scanning tunneling microscopy and/or spectroscopy (Galperin 2004).

Inelasticity in the electron transport through molecular junctions is closely related to the dephasing effects. One may say that incoherent electron transport always includes an inelastic contribution with the possible exception of the low temperature range. The general approach to theoretically analyze electron transport through molecular junctions in the presence of dissipative/phase-breaking processes in both electronic and nuclear degrees of freedom is based on the advanced formalisms ((Segal 2000, Skourtis 1995, Wingreen 1989,1993). These microscopic computational approaches have the advantages of being capable of providing the detailed dynamics information. However, while considering stationary electron transport through molecular junctions, one may turn to less time consuming approach based on the scattering matrix formalism (Buttiker 1986, Li 2001), as discussed below.

\subsection{II. Coherent transport}

To better show the effects of dissipation/dephasing on the electron transport through molecular junctions it seems reasonable to start from the case where these effects do not occur. So, we consider a molecule (presented as a set of energy levels) placed in between two electrodes with nearly continuous energy spectra. While there is no bias voltage applied across the junction, the latter remains in equilibrium characterized by the equilibrium Fermi energy $E_F,$ and there is no current flowing through it. When the bias voltage is applied, it keeps the left and right electrodes at different electrochemical potentials $\mu_L$ and $\mu_R.$ Then the electric current appears in the junction, and molecular energy levels located in between the electrochemical potentials $ \mu_L$ and $ \mu_R $ take the major part in maintaining this current. Electrons from occupied molecular states tunnel to the electrodes in accordance with the voltage polarity, and the electrons from one electrode travel to another one using unoccupied molecular levels as intermediate states for tunneling. Usually, the electron transport in molecular junctions occurres via highest occupied (HOMO) and lowest unoccupied (LUMO) molecular orbitals  which work as channels for electron transmission. Obviously, the current through the junction depends on quality of contacts between the leads and the molecule ends. However, there also exist the limit for the conductance in the channels. As was theoretically shown (Landauer 1970), the maximum conductance of a channel with a single spin degenerate energy level equals:
  \be 
 G_0 = \frac{e^2}{\pi\hbar} = (25.8k\Omega)^{-1} \label{1}
  \ee
 where $e$ is the electron charge and $ \hbar$ is the Planck's constant. This is a truly remarkable result for it proves that the minimum resistance $ R_0 = G_0^{-1} $ of a molecular junction cannot become zero. In another words one never can short-circuit a device operating with quantum channels. Also, the expression (\ref{1}) shows that the conductance is a quantized quantity. 

Conductance $g$ in practical quantum channels associated with molecular orbitals can take on values significantly smaller that $G_0,$ depending on the  delocalization in the molecular orbitals participating in the electron transport. In molecular junctions it also strongly depends on  the molecule coupling to the leads   (quality of contacts), as was remarked before. The total resistance $r = g^{-1} $ includes contributions from contact and molecular resistances, and could be written as (Wingreen 1993):
  \be 
 r = \frac{1}{G_0} \left(1 + \frac{1 - T}{T}\right).  \label{2}
 \ee
 Here, $ T $ is the electron transmission coefficient which generally takes on values less than unity.

The general expression for the electric current flowing through the molecular junction could be obtained if one calculates the total probability for an electron to travel between two electrodes at a certain tunnel energy $ E$ and then integrates the latter over the whole energy range (Datta 1995). This results in the well known Landauer expression:
  \be
 I = \frac{e}{\pi\hbar} \int T(E)\big[f_L(E) - f_R(E)\big]dE.  \label{3}
 \ee
 Here,  $f_{L,R}(E)$ are Fermi distribution functions for the electrodes with chemical potentials $\mu_{L,R},$ respectively. The values of $\mu_{L,R} $ differ from the equilibrium energy $E_F$, and they are determined with the voltage distribution inside the system. Assuming that the coherent tunneling predominates in the electron transport, the electron transmission function is given by (Datta 1997, Samanta 1996):
  \be 
 T(E) = 2 Tr \{\Delta_L G \Delta_R G^+ \}   \label{4}
 \ee
  where the matrices $ \Delta_{L,R} $ represent imaginary parts of self-energy terms $ \Sigma_{L,R}$ describing the coupling of the molecule to the electrodes, and $ G $ is the Green's function matrix for the molecule whose matrix elements between the molecular states $\big<i|$ and $|j\big>$ have the form:  
  \be
  G_{ij} = \big<i|E - H|j\big>. \label{5}
   \ee
 Here, $H $ is the molecular Hamiltonian including self-energy parts $ \Sigma_{L,R}.$

When a molecule contacts the surface of electrodes, this results in charge transfer between the molecule and the electrodes, and in modification of the molecule energy states due to redistribution of the electrostatic potential within the molecule. Besides, the external voltage applied across the junction brings additional changes  to the electrostatic potential further modifying molecular orbitals. The coupling of the molecule to the leads  may also depend on the voltage distribution. So, generally, the electron transmission $ T$ inserted in the Eq. (\ref{3}) and electrochemical potentials $\mu_{L,R}$ depend on the electrostatic potential distribution in the system.  To find the correct distribution of  the electric field inside the junction one must simultaneously solve the Schrodinger equation for the molecule and the Poisson equation for the charge density, following a self-consistent converging procedure. This is a nontrivial and complicated task, and significant effort was applied to study the effect of electrostatic potential distribution on the electron transport through molecules (Damle 2001, Di Ventra 2000, Galperin 2006, Lang 2000, Mujica 2000, Xue 2001,2003,2004). Here we put these detailed considerations aside, and we use the simplified expression for the electrochemical potentials:
  \be
 \mu_L = E_F + \eta|e| V; \qquad 
\mu_R = E_F - (1 - \eta) |e| V  \label{6}
 \ee
 where the parameter $ \eta $ indicates how the bias voltage is distributed between the electrodes. Also, we assume that inside the molecule the external electrostatic field is screened due to the charge redistribution, and the electron transmission is not sensitive to the changes in the voltage $V.$ Although very simple, this model allows to analyze the main characteristics of the electron transport through molecular junctions. Within this model one may write down the following expression for the self-energy parts ((D'Amato 1990):
   \be
 (\Sigma_\beta)_{ij} = \sum_k \frac{\tau_{ik,\beta}^* \tau_{kj,\beta}}{E - \epsilon_{k,\beta} + is}.  \label{7}
  \ee
 Here, $\beta \in L,R, \ \tau_{ik,\beta} $ is the coupling strength between $``i"$-th molecular state and $``k"$-th state on the left/right lead, $\epsilon_{k,\beta}$ are the energy levels on the electrodes, and $ s $ is a positive infinitesimal parameter. Assuming that the molecule is reduced to a single orbital with the energy $E_0$ (a single-site bridge), the Green's function accepts the form:
  \be
 G(E) = \frac{1}{E -E_0 - \Sigma_L - \Sigma_R}. \label{8}
 \ee
 Accordingly, in this case one may simplify the expression (\ref{4}) for the electron transmission: 
  \be
 T(E) = \frac{4 \Delta_L \Delta_R}{(E -E_0)^2 + (\Delta_L + \Delta_R)^2} .  \label{9}
 \ee

\begin{figure}[t] 
\begin{center}
\includegraphics[width=4.7cm,height=9cm,angle=-90]{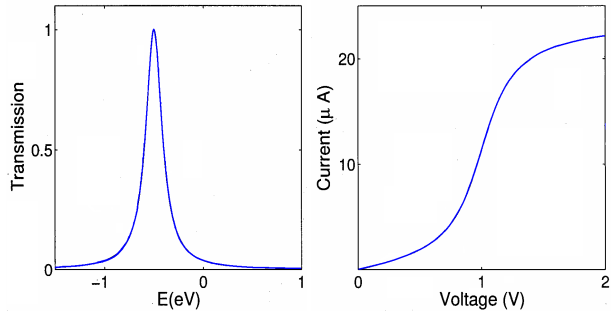}
\caption{(Color online) Coherent electron transmission (left panel) and current (right panel) versus bias voltage applied across a molecular junction where the molecule is simulated by a single electronic state. The curves are plotted assuming $\Delta_L = \Delta_R = 0.1 eV,\ E_0 = -0.5 eV,\ T=30K.$
}
 \label{rateI}
\end{center}\end{figure}

To elucidate the main features of electron transport through molecular junctions we consider a few examples. In the first example we mimic a molecule as a one-dimensional chain consisting of $ N $ identical hydrogen-like atoms with nearest neighbors interaction. We assume that there is one state per isolated atom with the energy $ E_0, $ and that the coupling  between the neighboring sites in the chain is characterized by the parameter $ b. $ Such a model was theoretically analyzed by D'Amato and Pastawski  and in some other works (see e.g. Mujica 1994). Basing on the Eqs. (\ref{4}), (\ref{5}), it could be shown that for a single-site chain $(N = 1) $ the electron transmission reveals a well distinguished peak at $ E = E_0, $, shown in the Fig. 2 The height of this peak is determined with the coupling of the bridge site to the electrodes. The peak in the electron transmission arises because the molecular orbital $E=E_0 $ works  as the channel/bridge for electron transport between the leads. Similar peaks appear in the conductance $ g = dI/dV.$ Assuming the symmetric voltage distribution $(\eta= 1/2)$ the peak in the conductance is located at $ V= \pm 2E_0.$ As for the current voltage characteristics, they display step-like shapes with the steps at $ V = \pm 2 E_0. $ When the chain includes several sites we obtain a set of states (orbitals) for our bridge instead of the single state $ E= E_0, $ and their number equals the number of sites in the chain. All these states are the channels for the electron transport. Correspondingly, the transmission reveals a set of peaks as presented in the Fig. 3. The peaks are located within the energy range with the width $4 b$ around $E = E_0.$ The coupling of the chain ends to the electrodes affects the transmission, especially near $ E =E_0. $ As the coupling strengthens, the transmission minimum values increase. Now, the current voltage curves exhibit a sequence of steps. The longer is the chain the more energy levels it possesses, and the more steps appear in the $I-V$ curves.

\begin{figure}[t] 
\begin{center}
\includegraphics[width=4.7cm,height=9cm,angle=-90]{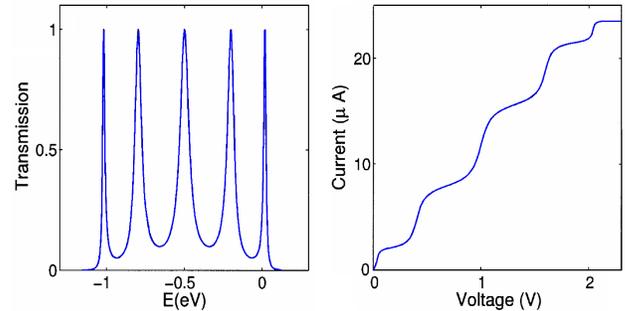}
\caption{(Color online)  Coherent electron transmission (left panel) and current (right panel) through a junction with a five electronic states bridge. The curves are plotted for $\Delta_L = \Delta_R = 0.1 eV,\ b = 0.3 eV,\ E_0 = - 0.5 eV,\ T =30K.$
}
 \label{rateI}
\end{center}\end{figure}

The second example concerns the electron transport through a carbon chain placed between copper electrodes. In this case, as well as for practical molecules a preliminary step in transport calculations is to compute the relevant molecular energy levels and wave functions. Usually, these computations are carried out employing quantum chemistry software packages (e.g. GAUSSIAN) or density-functional based software. Also, a proper treatment of the molecular coupling to the electrodes is necessary for it brings changes into the molecular energy states. For this purpose one may use the concept of an ``extended molecule'' proposed by Xue, Datta and Ratner (Xue 2001). The point of this concept is that only a few atoms on the surface of the metallic electrode are significantly disturbed when the molecule is attached to the latter. These atoms are located in the immediate vicinity of the molecule end. Therefore one may form a system consisting of the molecule itself and the atoms from the electrode surfaces perturbed by the molecular presence. This system is called the extended molecule and treated as such while computing the molecular orbitals. In the considered example the extended molecule included four copper atoms at the each side of the carbon chain. The results for electron transport are shown in the Fig. 4. Again, we observe a comb-like shape of the electron transmission corresponding to the set of transport channels provided by the molecular orbitals, and the stepwise $I-V$ curve originating from the latter. Transport calculations similar to those described above were repeatedly carried out in the last two decades for various practical molecules (see e.g. Xue 2001,2003, Galperin 2006, Zimbovskaya 2002,2003,2008). 
\begin{figure}[t]  
\begin{center}
\includegraphics[width=5cm,height=10cm,angle=-90]{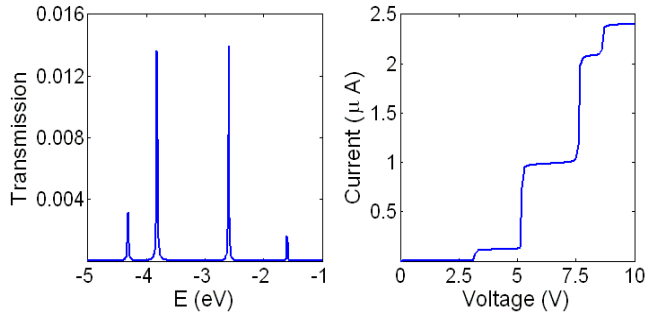}
\caption{(Color online)  Coherent electron transmission (left panel) and current (right panel) through a carbon chain coupled to the copper leads at $ T = 30 K.$
}
 \label{rateI}
\end{center}\end{figure}

\subsection{III. Buttiker model for inelastic transport} 

An important advantage of the phenomenological model for the incoherent/inelastic quantum transport proposed by Buttiker (Buttiker 1986) is that this model could be easily adapted to analyze various inelastic effects in the electron transport through molecules (and some other mesoscopic systems) avoiding complicated and time consuming advanced methods such as based on the non-equilibrium Green's functions formalism (NEGF).

Here, we present the Buttiker model for a simple junction including two electrodes linked by a single-site molecular bridge. The bridge is attached to a phase-randomizing electron reservoir, as shown in the Fig. 5. Electrons tunnel from the electrodes to the bridge and vice versa via the channels $1$ and $2$. While on the bridge, an electron could be scattered into the channels $3$ and $4$ with a certain probability $\epsilon.$ Such electron arrives at the reservoir where it undergoes inelastic scattering accompanied by phase-breaking and then  the reservoir reemits it back to the channels $3$ and $4$ with the same probability. So, within the Buttiker model the electron transport through the junction is treated as the combination of tunnelings through the barriers separating the molecule from the electrodes and interaction with the phase-breaking electron reservoir coupled to the bridge site. The key parameter of the model is the probability $\epsilon$ which is closely related to the coupling strength between the bridge site and the reservoir. When $\epsilon = 0$ the reservoir is detached from the bridge, and the electron transport is completely coherent and elastic. Within the opposite limit $(\epsilon = 1)$ electrons are certainly scattered into the reservoir which results in the overall phase randomization and inelastic transport.
\begin{figure}[t] 
\begin{center}
\includegraphics[width=5.2cm,height=7cm,angle=-90]{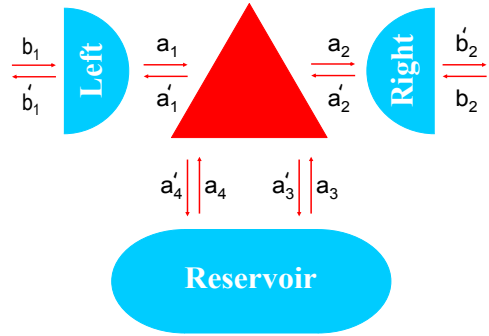}
\caption{(Color online)  Schematic drawing illustrating inelastic electron transport through a molecular junction within the Buttiker model.
}
 \label{rateI}
\end{center}\end{figure}

Within the Buttiker model the particle fluxes outgoing from the junctions  $J_i'$ could be presented as linear combinations of the incoming fluxes $J_k$ where the indexes $i,k$ label the channels for the transport: $ 1 \leq i,\ k \leq 4. $ 
   \be
 J_i' = \sum_k T_{ik} J_k.   \label{10}
  \ee
 The coefficients $ T_{ik} $ in these linear combinations are matrix elements of the transmission matrix which are related to the elements of the scattering matrix $ S,$ namely: $ T_{ik} = |S_{ik}|^2. $
 The  matrix $S$  expresses outgoing wave amplitudes $b_1',b_2',a_3',a_4'$ in terms of the incident ones $b_1,b_2,a_3,a_4.$ 
 To provide the charge conservation in the system, the net current in the channels $3$ and $4$ linking the system with the dephasing reservoir must be zero, so we may write:
  \be 
  J_3 + J_4 - J_3' -J_4' = 0 .  \label{11}
  \ee
 The transmission for the quantum transport could be defined as the ratio of the particle flux outgoing from the system and that one incoming to the latter. Solving Eqs. (\ref{10}), (\ref{11}) we obtain:
  \be
 T(E) = \frac{J_2'}{J_1} = T_{21} + \frac{K_1 \cdot K_2}{2 – R}  \label{12}
 \ee
 where:
 \begin{align}
& K_1 = T_{31} + T_{41};  
  \qquad
K_2 = T_{23} + T_{24};
 \nn\\
 & R = T_{33} + T_{44} + T_{43} + T_{34}.   \label{13} 
\end{align}
  For the junction including the single-site bridge the scattering matrix $ S $ has the form (Buttiker 1986): 
  \be
  S = \frac{1}{Z}   \left(
\begin{array}{cccc}%
r_1 + \alpha^2 r_2 & \alpha t_1 t_2 & \beta t_1 & \alpha \beta t_1 r_2  
  \\
\alpha t_1 t_2 & r_2 + \alpha^2 r_1 & \alpha \beta r_1 t_2 & \beta t_2 
   \\
\beta t_1 & \alpha \beta r_1 t_2 & \beta^2 r_1 & \alpha r_1 r_2 - \alpha
   \\
\alpha \beta t_1 r_2 & \beta t_2 & \alpha r_1 r_2 - \alpha & \beta^2 r_2
\end{array}
\right)  . \label{14}
  \ee 
  Here, $ Z = 1 - \alpha^2 r_1 r_2 , \ \alpha = \sqrt{1 - \epsilon},\ \beta = \sqrt\epsilon $ and $ r_{1,2} $ and $t_{1,2} $ are the amplitude reflection and transmission coefficients for the two barriers.
 Later the expression for this matrix suitable for the case of multi-site bridges including several inelastic scatterers was derived (Li 2001).

Assuming for certainty the charge flow from the  left to the right we may write down the following expression (Zimbovskaya 2005):
  \be
 T(E)  = \frac{g(E)(1 +\alpha^2)[g(E)(1 + \alpha^2) + 1 - \alpha^2\big]}{\big[g(E)(1 - \alpha^2) + 1 + \alpha^2 \big]^2} \label{15}
  \ee  
  where 
  \be
 g(E) = 2 \sqrt{\frac{\Delta_L \Delta_R}{(E - E_0)^2 + (\Delta_L + \Delta_R)^2}} .               \label{16}
        \ee
 Now, the electron transmission strongly depends on the dephasing strength $\epsilon. $ As shown in the Fig. 6 coherent transmission $(\epsilon = 0)$  exhibits a sharp peak at $E =E_0$ which gives a step-like shape to the volt-ampere curve, as was discussed in the previous section.  In the presence of dephasing the peak gets eroded. When the $\epsilon$ value approaches $1$  the $I-V$ curve becomes linear, corroborating the ohmic law for the inelastic transport.

\begin{figure}[t] 
\begin{center}
\includegraphics[width=4.7cm,height=9.2cm,angle=-90]{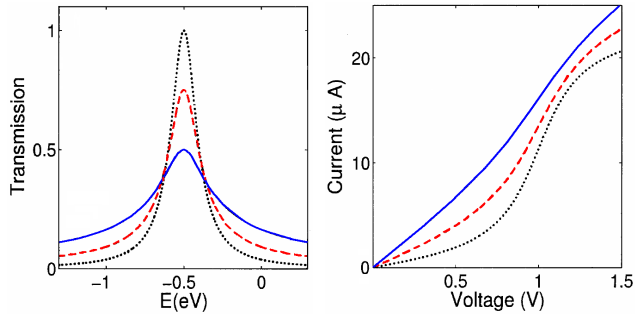}
\caption{(Color online)  Electron transmission (left panel) and current (right panel) computed within the Buttiker model at various values of the dephasing parameter $\epsilon,$ namely: $\epsilon = 0 $ (dotted lines), $ \epsilon = 0.5 $ (dashed lines), and $ \epsilon = 1 $ (solid lines). The curves are plotted assuming that the molecule is simulated by a single orbital with $E_0 = - 0.5 eV$ at $\Delta_L = \Delta_R = 0.1 eV,\ T = 30 K.$
}
 \label{rateI}
\end{center}\end{figure}
Within the Buttiker's model $\epsilon$ is introduced as a phenomenological parameter whose relation to the microscopic characteristics of the dissipative processes affecting electron transport through molecular junctions remains uncertain. To further advance this model one should find out how to express $\epsilon$ in terms of the relevant microscopic characteristics for various transport mechanisms. This should open the way to make a link between the phenomenological Buttiker model and NEGF. Such attempt was carried out in recent works (Zimbovskaya 2005,2008) where the effect of stochastic nuclear motion on the electron transport through molecules was analyzed.

\subsection{IV. Vibration-induced inelastic effects}

Interaction of electrons with molecular vibrations is known to be an important source of inelastic contribution to the electron transport through molecules. Theoretical studies of vibrationally inelastic electron transport through molecules and other similar nanosystems (e.g. carbon nanotubes) were carried out over the past few years by a large number of authors (Cornaglia 2004,2005, Donarini 2006, Egger 2008, Galperin 2007,Gutierrez 2006, Kushmerick 2004, Mii 2003, Rynduk 2007, Siddiqui 2007, Tikhodeev 2004, Troisi 2006, Zazunov 2006, 2007,  Zimmerman 2008). Also, manifestations of the electron-vibron interactions were experimentally observed (Agrait 2003, Djukic 2005, Lorente 2001, Qiu 2004, Repp 2005,  Segal 2001, Smit 2004, Tsutsui 2006, Wang 2004, Wu 2004, Zhitenev 2002). To analyze vibration induced effects on the electron transport through molecular bridges, one must assume that molecular orbitals are coupled to the phonons describing  vibrations. While on the bridge, electrons may participate in the events generated by their interactions with vibrational phonons. These events involve virtual phonon emission and absorption. For rather strong electron-phonon interaction this leads to the appearance of metastable electron levels which could participate in the electron transport through the junctions, bringing an inelastic component to the current. As a result, vibration induced features occur in the differential molecular conductance $dI/dV$ and in the inelastic tunneling spectrum $d^2I/dV^2 $. This was observed in the experiments (see e.g. Qiu 2004, Zhitenev 2002). Sometimes even current voltage curves themselves exhibit extra step originating from the electron-vibron interactions (Djukic 2005).

Particular  manifestations of electron-vibrionic effects in the transport characteristics are determined by the relation of three relevant energies. These are the coupling strengths of the molecule to the electrodes $ \Delta_{L,R},$  the electron-phonon coupling strength $\lambda$ and the thermal energy $kT\ (k$ is the Boltzman constant). When the molecule is weakly coupled to the electrodes $(\Delta_{L,R} \ll \lambda,\hbar\Omega)$ and the temperature is low $(kT \ll \Delta_{L,R}),$ the electron transfer through the junction may give rise to a strong vibrational excitation, and one may expect pronounced vibrational resonance structure in the electron transmission to appear. Correspondingly, extra steps should occur in the $I-V$ curves. Proper theoretical consideration of the electron transport in this regime could be carried out within the approach proposed by Wingreen {\it et al} (Wingreen 1989). Here, we employ a very simple semi-quantitative approximation which, nevertheless, allows to qualitatively describe this structure including the effect of higher phonon harmonics on the transport characteristics.
\begin{figure}[t] 
\begin{center}
\includegraphics[width=4.5cm,height=9cm,angle=-90]{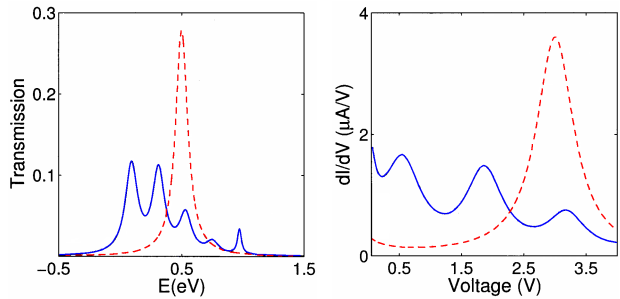}
\caption{(Color online) The electron transmission versus energy (left panel) and the conductance versus voltage (right panel) for a junction with the  molecular bridge simulated by a single electronic state weakly coupled to the leads: $ \Delta_L = \Delta_R = 0.01 eV,\ E_0 = 0.5 eV.$ Solid lines are plotted assuming that the bridge is coupled to a single phonon mode $(\hbar \Omega = 0.22 eV,\ \lambda = 0.3 eV).$ Dashed lines correspond to the coherent electron transport. 
}
 \label{rateI}
\end{center}\end{figure}

We consider a junction including a single-site bridge which is coupled to a single vibrational mode with the frequency $\Omega.$ An electron on the bridge may virtually absorb several phonons which results in the creation of a set of metastable states with the energies $E_n = \tilde E_0 + n \hbar \Omega\ (n = 0,1,2,...).$ Here, the energy $\tilde E_0$ is shifted with respect to $E_0$ due to the electron-phonon interaction. The difference in these energies $E_p$ is called a polaronic shift and could be estimated as $E_p = \lambda^2/\hbar\Omega$ (Gutierrez 2006, Ryndyk 2007, Wingreen 1989). At weak coupling of the bridge state to the electrodes the lifetime of these metastable states is long enough for them to serve as channels for the electron transmission. Therefore, one may roughly approximate the transmission as a sum of contributions from all these channels. The terms in the sum have the form similar to the well-known expression for the coherent transmission (see Eq. (\ref{9})). However, every term includes the factor $P(n)$ which corresponds to the probability of the metastable state to appear. So, we obtain:
  \be
 T(E) = 2 \Delta_L\Delta_R \sum_n \frac{P(n)}{(E - \tilde E_0 - n\hbar\Omega)^2 + (\Delta_L + \Delta_R)^2} . \label{17}
  \ee
 Here (Cizek 2004):   
  \be
 P(n) = \frac{1}{n!} \left(\frac{\lambda^2}{2\hbar^2\Omega^2}\right)^n \exp 
\left(- \frac{\lambda^2}{2\hbar^2\Omega^2}\right)    .  \label{18}
 \ee
 The phonon induced peaks in the transmission are displayed in the Fig. 7 along with the transmission peak for the coherent transport through a single-site bridge. As expected, the coupling of electronic degrees of freedom to the vibrational motion  splits the single peak in the coherent transmission into the set of smaller peaks associated with vibrational levels. The peaks could be resolved when $ \Delta_{L,R} \ll \hbar\Omega.$ This agrees with the results of the earlier theoretical works (Wingreen 1989) as well as with the experiments (Qiu 2004, Zhitenev 2002). Phonon-induced peaks in the transmission give rise to the steps in the $I-V$ curves and rather sharp features (peaks and dips) in the inelastic tunneling spectrum (IETS). The latter are shown in the Fig. 8 (left panel) and they resemble those obtained using proper NEGF based calculations (Galperin 2004).

\begin{figure}[t] 
\begin{center}
\includegraphics[width=4.9cm,height=9.8cm,angle=-90]{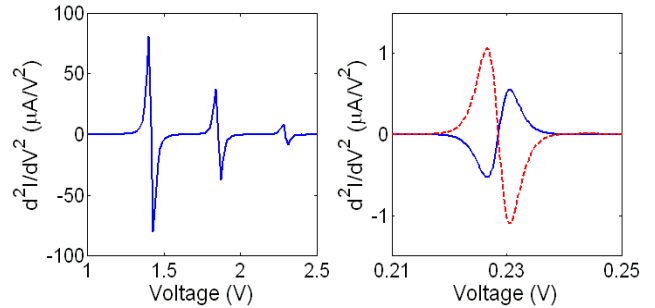}
\caption{(Color online) The inelastic electron tunneling spectrum plotted against the bias voltage at $\hbar \Omega = 0.22 eV,\ \lambda = 0.3 eV.$ Left panel:  $ \Delta_L = \Delta_R = 0.01 eV,\ E_0 = 0.5 eV.$  Right panel:  $ \Delta_L = \Delta_R = 0.5 eV,\ E_0 = 0.5 eV$ (solid line), $ E_0 = -0.5 eV $ (dashed line). 
}
 \label{rateI}
\end{center}\end{figure}

When the molecule is strongly coupled to the electrodes $(\Delta_{L,R} \gg \lambda)$ and the temperature is still low $(kT \ll \lambda,\hbar\Omega),$ electron-vibrionic interaction effects are less pronounced. Both current and conductance are weakly affected by the electron-phonon coupling (Galperin 2007, Tikhodeev 2004). However, the IETS features  remain distinguishable. These features appear at the threshold $V=\hbar\Omega/|e| $ which corresponds to the opening of a channel for inelastic transport.  To analyze IETS in a simplest way one may use the result for electron transmission derived within the Buttiker model where the dephasing parameter $\epsilon$ is expressed in terms of the relevant energies, namely:
  \be
 \epsilon = \frac{\Gamma_{ph}}{2(\Delta_L + \Delta_R) + \Gamma_{ph}} \label{19}
  \ee 
  where $ \Gamma_{ph} = 2 \mbox{Im} (\Sigma_{ph}),$ end $\Sigma_{ph} $ is the self-energy term originating from the  electron-phonon interaction. Basing on the nonequilibrium Green's function  formalism, the expression for $ \Gamma_{ph} $ was derived  in the form (Mii 2003):
   \begin{align}
 \Gamma_{ph} (E) = & \ 2\pi\lambda^2 \int d\omega \rho(\omega) \big\{N(\omega) 
[\rho_{el} (E - \hbar\omega) + \rho_{el}(E + \hbar\omega)] 
   \nn\\ &
 + (1 + N(\omega)) ([1 - n(E - \hbar\omega)] \rho_{el} (E -\hbar\omega)
   \nn\\
  & + n(E + \hbar\omega) \rho_{el}(E + \hbar\omega)) 
  \nn\\ &
+ N(\omega) 
\big([1 - n (E + \hbar\omega)] \rho_{el}(E + \hbar\omega) 
   \nn\\ & + 
n (E - \hbar\omega) \rho_{el}(E -  \hbar\omega) \big) \big \} .   \label{20}
   \end{align} 
  Here, $\rho_{el}(E)$ and $\rho_{ph}(\omega) $ are the phonon and electron densities of states, respectively; $ N(\omega)$ is the Bose-Einstein distribution function at the temperature $ T,$ and:
   \be
 n(E) = \frac{1}{2} \big[f_L(E) + f_R(E) \big].  \label{21}
  \ee
 While considering a junction with a single-site bridge state coupled to the single vibrational mode, we can write the following expressions for $\rho_{ph}$ and $\rho_{el}:$
  \be
 \rho_{ph}(\omega) = \frac{1}{\pi\hbar} \frac{\gamma}{(\omega - \Omega)^2 + \gamma^2} ,   \label{22}
  \ee
  \be
  \rho_{el}(E) = \frac{1}{\pi} \frac{\Delta_L + \Delta_R}{(E - E_0)^2 + (\Delta_L + \Delta_R)^2}     \label{23}
  \ee
  where the polaron shift is neglected, and the parameter $\gamma $ characterizes the broadening of the maximum in $\rho_{ph} $ at $ \omega = \Omega $ due to the interaction of the vibrionic mode with the environment. At low temperatures we may significantly simplify the expression for $\Gamma_{ph}.$ Within the conduction window $ \mu_R < E < \mu_L $ we get:
   \begin{align}
 \Gamma_{ph}(E) \approx &\pi\lambda^2 \left \{\int_0^{(\mu_L - E)/\hbar} d\omega \rho_{ph}(\omega) \rho_{el}(E + \hbar\omega) \right.
\nn \\ &   + \left.
 \int_0^{(E - \mu_R)/\hbar} d\omega \rho_{ph}(\omega) \rho_{el} (E - \hbar\omega) \right\}  . \label{24}
  \end{align}
  Omitting from consideration the coupling of the phonon mode to the environment $(\gamma \to 0)$ we may easily carry out integration  over $ \omega $ in the Eq. (\ref{24}), and we arrive at the result:
   \begin{align}
   \Gamma_{ph} (E) \approx & \pi\lambda^2 \big\{\rho_{el}(E + \hbar\Omega) \theta(\mu_L - \hbar\Omega - E) 
\nn \\ &
+ \rho_{el}(E - \hbar\omega) \theta(E - \mu_R - \hbar\Omega)\big\}  \label{25}
  \end{align}
 where $ \theta(x)$ is the step function.

Substituting the approximation for $ \Gamma_{ph} $ given by Eq. (\ref{24}) or Eq. (\ref{25}) into the expression (\ref{19}) for the dephasing parameter $ \epsilon$ and employing the earlier result (\ref{15}) for the electron transmission, we may calculate transport characteristics. The adopted simplified  approach, as well as NEGF based calculations, shows that the IETS signal appears at the threshold determined by the frequency of the vibrational mode (see Fig. 8, right panel). 
 At first glance one may expect the net current through the junction to increase at the threshold. Indeed, the inelastic contribution to the current increases from zero to a certain nonzero value at this threshold for the channel for inelastic transport opens up. However, more thorough studies show that both elastic and nonelastic contributions to the net current undergo changes at the inelastic tunneling threshold, and the elastic current could decrease there, as was first shown by Persson and Baratoff (Persson 1987). Moreover, this decrease in the elastic current may overweigh the contribution coming from the inelastic channel. Depending on the relative value of the elastic and inelastic contributions to the net current near threshold, the IETS reveals a peak or a dip at the corresponding voltage. Experiments corroborate the variety in the IETS taken for molecular junctions (Hahn 2000, Djukic 2005, Wang 2004, Zhitenev 2002). 
The shape of the signal is very sensitive to the characteristics of the junction such as the position of the electronic state, electron-phonon and molecule-to leads coupling strengths, and the vibrionic frequency.
For instance, at a very strong coupling of the molecular bridge to the leads, it could so happen that the backscattered by the negatively biased electrode electrons whose energies belong to the conduction window between $ \mu_L$ and and $\mu_R,$ are locally depleted near the junction. In this situation the opening of the inelastic channel may cause the increased reflection (otherwise forbidden by the Fermi exclusion) leading to the decrease in the conduction. Consequently, the described scenario  should result in the dip in IETS signal. Also, as concluded in the recent work of Ryndyk and Cuniberti (Ryndyk 2007) the above discussed sideband phonon-induced features in the electron spectral density could give rise to the corresponding features in the differential conductance $ dI/dV$ and IETS assuming that the molecule coupling to the leads is not too strong. Contributions from these sideband features may be responsible for the shape of the IETS signal at the threshold of the inelastic tunneling channel. These contributions could produce an extra inelastic signal, as well. The latter appears as an additional peak or dip in the differential conductance. 

The question of current decrease/increase at the phonon excitation threshold which corresponds to the peak/dip in the IETS is not completely answered so far, and the appropriate theory is being developed (Balseiro 2006, Egger 2008).      
Nevertheless, it is presently understood that three relevant energies, namely, molecule-electrode couplings $\Delta_{L,R}, $ electron-phonon coupling strength $ \lambda$ and the phonon energy $ \hbar\Omega $ play very important part in determining the shape of the IETS signal.    
Varying these parameters one may convert the IETS signal from a peak to a dip and vice versa. 

At finite temperatures molecular vibrations always occur in the presence of stochastic nuclear motions. These motions could be described as a phonon thermal bath. Coupling of vibrational modes to this bath further affects the electron transport causing energy dissipation. The dissipative processes must be taken into account to properly analyze the effects of electron-phonon interactions in the electron transport. Also, the displacements of ions involved in the molecular vibrations are accompanied with the changes in the electrostatic field inside the molecule. This could give rise to polaronic effects in the electron transport. We discuss these issues in the next sections.

\subsection{V. Dissipative transport}

Electron transport through molecular junctions is always accompanied by stochastic nuclear/ion motions in the close environment. Interactions of travelling electron with these environmental fluctuations causes energy dissipation. The importance of dissipative effects depends on several factors. Among these factors the temperature and the size and complexity of the molecular bridge are predominating. The temperature determines the intensity of the nuclear motions, and the size of the molecule determines so-called contact time, that is the time for electron to travel through the junction and, in consequence, to contact the environment. It was shown by Buttiker and Landauer (Buttiker 1985) that the contact time is proportional to the number of sites (subunits) in the molecule providing intermediate states for the electron tunneling. For small molecules at low temperatures the contact time is shorter than characteristic times for fluctuations in the environment, so, the effect of the latter on the electron transport is not very significant. One may expect small broadening of the molecule energy states to occur which brings a moderate erosion of the peaks in the electron transmission and steps in the $ I-V$ characteristics.           

On the contrary, in large sized molecules such as proteins and DNA, electron transport is accompanied by strong energy dissipation. The significance of the system-environment interactions in macromolecules was recognized long ago in studies of long range electron transfer reactions. In these reactions electrons travel between distant sites on the molecule called a donor and an acceptor. It was established that when an electron initially localized on the donor site moves to the acceptor site with a lower energy, the energy difference must be dissipated to the environment to provide the irreversibility of the transfer (Garg 1985).

A usual way to theoretically analyze dissipative effects in the electron transport through molecules is to introduce a phonon bath  representing the random motions in the environment. In general, there is no one/several dominating modes in the bath. Instead, the bath is characterized by a continuous spectral function of all relevant phonon modes $\rho_{ph}(\omega).$ The electrons are supposed to be coupled to the phonon bath, and this coupling is specified   
by the spectral function. The particular form of $\rho_{ph}(\omega)$ may be found basing on the molecular dynamics simulations. However, to qualitatively study the effect of dissipation on the electron transport one can employ the expression (Mahan 2000):
  \be
 \rho_{ph}(\omega) = \lambda \frac{\omega}{\omega_c} \exp\left(-\frac{\omega}{\omega_c} \right)
   \label{26}  \ee
 where the parameter $ \lambda $ characterizes electron-phonon coupling  strength, and $\omega_c$ is the cutoff frequency for the bath related to its thermal relaxation time $ \tau_c = \omega_c^{-1}.$

To illustrate possible effects  of dissipation on the electron transport we return back to our simple model where the molecular bridge is represented by a single state. Now, we assume that this state is coupled to the phonon bath. This model is hardly appropriate to properly analyze dissipative effects in the electron transport through practical molecules for  the molecule length is very important for dissipative effects to be pronounced. Nevertheless, it still could serve to basically outline main features of the dissipative electron transport through molecular junctions. Also, the proposed model could be useful to analyze electron transport in doped polyacetylene/polyaniline-polyethylene oxide nanofibers (Zimbovskaya 2008). These conducting polymers could be treated as some kind of granular metals where metallic-like regions (grains) are embedded in the poorly conducting medium of disorderly arranged polymer chains (MacDiarmid 2001). While in metallic state, the intergrain electron transport in these nanofibers is mostly provided by electron tunneling through intermediate states on the polymer chains between the grains (Prigodin 2002). In this case the contact time could be long enough for the effects of dissipation to be well manifested which justifies the adoption of the above model.

Again, one may carry out transport calculations using Eq. (\ref{15}) for the electron transmission, and expressing $\epsilon $ in terms of of the relevant energies.  Substituting Eq. (\ref{26}) in the expression (\ref{20}) we may calculate $\Gamma_{ph}(E).$ The energy dissipation effects are more distinctly prononced at moderately high temperatures, so we assume $ kT \gg \hbar\omega_c. $  Then the main contribution to the integral over $ \omega $ in the Eq. (\ref{20}) originates from the low frequency region $ \omega \ll \omega_c;$ and we obtain the following approximation:
  \be
 \Gamma_{ph}(E) = \frac{8kT \lambda \Gamma}{(E - E_0)^2 + \Gamma^2}            
   \label{27}    \ee
  where $\Gamma = \Delta_L + \Delta_R + \frac{1}{2} \Gamma_{ph}(E).$ Solving the obtained equation for $ \Gamma_{ph} $ and using the Eq. (\ref{19}) we obtain (Zimbovskaya 2008):
  \be 
 \epsilon = \frac{1}{2} \frac{\rho^2\big(1 + \sqrt{1 + \rho^2}\big)}{\ds4 \Big(\frac{E-E_0}{\Delta_L + \Delta_R}\Big)^2 + \frac{1}{2}\big(1 + \sqrt{1 + \rho^2}\big)^3}   \label{28}
 \ee
 where $\rho^2 = 32kT\lambda/(\Delta_L + \Delta_R)^2. $ 

Voltage dependencies of the conductance computed basing on Eqs. (\ref{3}), (\ref{15}), (\ref{28}) are presented in the Fig. 10 (left panel). One may see that at low values of the bias voltage the electrons coupling to the phonon bath brings an enhancement in the conduction. The effect becomes reversed as the voltage grows above a certain value. This happens because the phonon-induced broadening of the molecular level (the bridge) assists the electron transport at small bias voltage. When the voltage rises this effect is surpassed by the scattering effect of phonons  
which resists the electron transport. The significance of the electron-phonon interactions is determined by the ratio of the coupling constant $\lambda$ and the self-energy terms describing the bridge coupling to the electrodes. The phonon bath makes a significant effect on the transport characteristics when $\lambda > \Delta_{L,R}.$

A dissipative electron transport through large DNA molecules was  studied both theoretically and experimentally (Gutierrez 2005, Xu 2004). Theoretical studies were based on a model where the molecule was simulated by a tight-binding chain of sites linking the electrodes, and attached side chains. Electrons are allowed to travel along the bridge chain and to hop to the near by side chains. These chains are coupled to the phonon bath providing the energy dissipation (see Fig. 9). Although proposed for the specific kind of poly$(d G)$-poly$(d C)$ molecules, this model seems to be quite generic and useful for a larger class of macromolecules.

\begin{figure}[t] 
\begin{center}
\includegraphics[width=4.7cm,height=7cm,angle=-90]{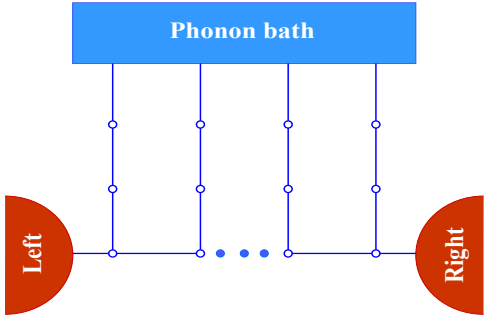}
\caption{(Color online)  Schematic drawing of a molecular junction where the molecular bridge is coupled to the phonon bath via side chains.
}
 \label{rateI}
\end{center}\end{figure}
Several coupling regimes to the bath may be analyzed. The most preferred regime for dissipative effects to appear is the strong-coupling limit defined by the condition $\lambda/\omega_c > 1.$ Within this regime the characteristic time for the electron bath interactions is much shorter than typical electron time scales. Consequently, the bath makes a significant impact on the molecule electronic structure. New bath- induced states appear in the molecular spectrum inside the HOMO-LUMO gap. However, these states are strongly damped due to the dissipative action of the bath (Gutierrez 2005). As a result, a small finite density of phonon-induced states appears inside the gap supporting electron transport at low bias voltage. So again, the environment induces incoherent phonon assisted transport through molecular bridges.
 For illustration we show here the results of calculations carried out for a toy model with a single-site bridge with a side chain attached to the latter. The side chain is supposed to be coupled to the phonon bath. The results for the electron transmission are displayed in the Fig. 10 (right panel). We see that the original bridge state at $E= 0$ is completely damped but two new phonon-induced states emerge nearby which could support electron transport.

\begin{figure}[t] 
\begin{center}
\includegraphics[width=4.7cm,height=9.2cm,angle=-90]{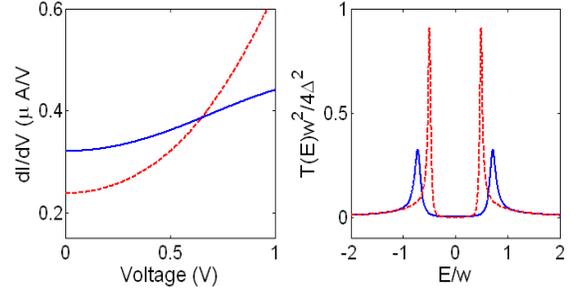}
\caption{(Color online) Left panel: The electron conductance versus voltage for a junction with a single electronic state  bridge directly coupled to the phonon thermal bath. The curves are plotted assuming $ E_0 = 0.4 eV,\ \Delta_L=\Delta_R = 0.1 eV,\ T = 30K,\ \lambda = 0.3 eV $ (solid (line), $ \lambda = 0.05 eV $ (dashed line). Right panel: Electron transmission through the junction in the case when the bridge state interacts with the phonon bath via the side chain coupled to the bridge state with the coupling parameter  $w.$ The curves are plotted assuming $ \Delta_L = \Delta_R = \Delta,\ w/\Delta = 20, \ E_0 = 0,\ \lambda = 0.3 eV $ (solid line), $ \lambda = 0.05 eV$ (dashed line).
}
 \label{rateI}
\end{center}\end{figure}
An important characteristics of the dissipative electron transport through  molecular junctions is the power loss in the junction, that is the energy flux from electronic into phononic system. Assuming the current flow from the left to the right, this quantity may be estimated as the sum of the energy fluxes $Q_{L,R}$ at the left and right terminals (leads): 
  \be 
 P = Q_L + Q_R.  \label{29}
 \ee
 One may express the energy fluxes  in terms of renormalized  currents at the electrodes $\tilde I_{L,R} (E)$ which are defined as follows (Datta 2005):
  \be
 I_{L,R} = \frac{e}{\pi\hbar} \int \tilde I_{L,R}(E) dE.  \label{30}
 \ee  
 Then $Q_{L,R}$ may be presented in the form:
  \be
 Q_{L,R} = \frac{1}{\pi\hbar} \int E\tilde I_{L,R}(E) dE.  \label{31}
 \ee
 We remark that the current $ I_{L,R}$ in the Eq. (\ref{30}) are related to the corresponding leads, and their signes are accordingly defined. An outgoing current is supposed to be positive whereas an incoming one is negative for each lead. To provide electric charge conservation in the junction one must require that $ I =I_L = -I_R$ for the chosen direction of the current flow. Therefore the energy fluxes also have different signs. As for the $Q_{L,R}$ magnitudes, they may differ only if the renormalized currents $\tilde I_L(E)$ and $\tilde I_R(E)$ are distributed over energies in different ways. This cannot happen in the case of elastic transport for in this case $\tilde I_L(E) = - \tilde I_R(E).$ However, if the transport process is accompanied with the energy dissipation, the energy distributions for $\tilde I_L $ and $ \tilde I_R $ may differ. In this case electrons lose some energy while moving through the junction and this gives rise to the differencies in the renormalized currents energy distributions. For instance, in the case when the electrodes are linked with a single-state bridge, the energy distribution of $ \tilde I_L(E)$ has a single maximum whose position is determined by the site energy $ E_0$ and the applied bias voltage $V.$ Assuming that the average energy loss due to dissipation could be estimated as $ \Delta E,$ the maximum in the $\tilde I_R(E)$ distribution is shifted by this quantity, so the current $  \tilde I_R(E)$ flows at lower energies compared to $\tilde I_L(E)$. This results in the power loss and Joul heating in the junction (Segal 2003). 

\subsection{VI. Polaron effects: Hysteresis, switching and negative differential resistance}
  
While studying electron transport through molecular junctions hysteresis in the current-voltage characteristics was reported in some systems (Li 2003). Multistability and stochastic switching  were reported in single-molecule junctions (Lortscher 2007) and in single metal atoms coupled to a metallic substrate through a thin ionic insulating film (Olsson 2007, Repp 2004).

Coupling of an electron belonging to a certain atomic energy level to the displacements of ions in the film brings a possibility of polaron formation in there. This leads to the polaron shift in the electron energy. 
It was noticed that multistability and hysteresis in molecular junctions mostly occurred when the molecular bridges included centers of long-living charged electronic states (redox centers). On these grounds it was suggested that hysteresis in the $I-V$ curves observed in  molecular junctions appear due to formation of polarons on the molecules (Galperin 2005).

The presence of the polaron shift in the energy of a charged (occupied) electron state creates a difference between the latter and the energy of the same state while it remains unoccupied. Assuming for simplicity a single-state model for the molecular bridge coupled to a single optical phonon mode we may write the following expression for the renormalized energy:
  \be
 \tilde E_0 (n_0) = E_0 - \frac{\lambda^2 n_0}{\hbar\Omega}   \label{32}
  \ee
 where the  electronic population on the bridge $n_0$ is given by: 
  \be
 n_0 = \frac{1}{\pi} \int dE \frac{f_L(E)\Delta_L + f_R (E)\Delta_R}{\big[E 
- \tilde E_0(n_0)\big]^2 + (\Delta_L + \Delta_R)^2} .    \label{33}
  \ee
 So, as follows from Eq. (\ref{32}) the polaron shift depends on the bridge occupation $ n_0,$ and the latter is related to $ \tilde E_0 $ by the Eq. (\ref{33}). Therefore, the derivation of an explicit expression for $ \tilde E_0(n_0)$ is a nontrivial task even within the chosen simple model. Nevertheless, it could be shown that two local minima emerge in the dependence of potential energy of the molecular junction including two electrodes linked by the molecular bridge, of the occupation number $n_0$. These minima are located near $ n_0 = 0 $ and $ n_0 = 1 ,$ and they correspond to the neutral (unoccupied)and charged (occupied) states, respectively. This is illustrated in the Fig. 11 (left panel). These states are metastable, and their lifetime could be limited by the quantum switching (Mitra 2005, Mozyrsky 2006). When the switching time between the two states is longer than the characteristic time for the external voltage sweeping, one may expect the hysteresis to appear for the states of interest live long enough to maintain it. Within the opposite limit the average washes out the hysteresis. When the states are especially short-lived this could even result in a telegraph noise at finite bias voltage which replaces the controlled switching.

\begin{figure}[t]  
\begin{center}
\includegraphics[width=4.7cm,height=9.2cm,angle=-90]{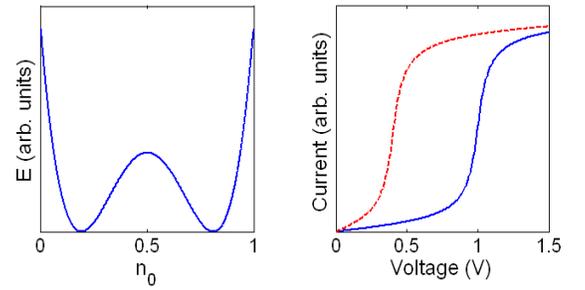}
\caption{(Color online) Left panel: Schematic of the potential energy of the molecular junction versus the occupation number $n_0.$ Right panel: Hysteresis in the current-voltage characteristics. The solid line corresponds to the transport via the unoccupied bridge electronic state,  and the dashed line corresponds to the transport via the occupied state shifted due to the polaron formation.
}
 \label{rateI}
\end{center}\end{figure}

Further we consider long-lived metastable states and we concentrate on the $ I-V$ behavior. Let us for certainty assume that the bridge state at zero bias voltage is situated above the Fermi energy of the system and remains empty. As the voltage increases, one of the electrodes chemical potentials crosses the bridge level position, and the current starts to flow through the system. The $ I-V$ curve reveals a step at the bias voltage value corresponding to the crossing of the unoccupied bridge level with the energy $ E_0$ by the chemical potential. However, while the current flows through the bridge, the level becomes occupied and, consequently, shifted due to the polaron formation. When the bias voltage is reversed  the current continue to flow through the shifted bridge state of the energy $ \tilde E_0, $ until the recrossing happens. Due to the difference in the energies of neutral and charged states the step in the $ I-V$ curves appears at different values of the voltage, and this is the reason for the hysteresis loop to appear as shown in the right panel of the figure 11. One could also trace the hysteresis in the $ I-V$ characteristics starting from the filled (and shifted) bridge state.   

Again, the hysteresis loop in the $ I-V$ curves may occur when both occupied and unoccupied states are rather stable which means that the potential barrier separating the corresponding minima in the potential energy profile is high enough, so that quantum switching between the states is unlikely. This happens when the bridge is weakly coupled to the electrodes $( \Delta_{L,R} \ll \lambda^2/\hbar\Omega) $ which is an obvious requirement for the involved states (neutral or charged) to the distinguishable. In other words, the broadening of the relevant levels due to the coupling to the electrodes must be much smaller that the polaron shift. As was recently shown (Ryndyk 2008), the stronger is the electron-phonon coupling the less becomes the probability of the switching. At large values of $ \lambda$ the tunneling between the charged and neutral states is exponentially suppressed. Also, it  was shown that the symmetry/asymmetry in the coupling to the electrodes could significantly affect the hysteresis behavior (D'Amico 2008, Ryndyk 2008). For asymmetric junctions $(\Delta_L \neq \Delta_R)$  two nearly stable states exist at zero bias voltage. When the asymmetry is very strong both states could appear stable at one bias voltage polarity and unstable when the polarity is reversed. It was suggested that under appropriate choice of parameters one may create a situation when the instability regions for the involved states do not overlap. These properties give grounds to conjecture that such strongly asymmetric junctions could reveal memory functionalities, which make them potentially useful in fabrication of nanodevices.   

 Among various potentially important properties of the electron transport through metal-molecule junctions one may separate out the negative differential resistance (NDR), that is the decrease in the current $ I $ while the bias voltage across the molecule increases. The NDR effect was originally observed in tunneling semiconducting  diodes (Kastner 1992). Later, it was viewed in quantum dots (Grobis 2005, Repp 2005). Several possible scenarios were proposed to explain the NDR occurrence in the electron transport through molecules. The effect could originate from alignment and subsequent disalignment  of the Fermi levels of the electrodes with molecular orbitals which happens as the bias voltage varies (Xue 1999). Also, the DNR could appear as a Coulomb blockade induced effect (Simonian 2007) and/or due to some other reasons. It is likely that different mechanisms could play a major part in the NDR appearance in different molecular junctions where it was observed so far.

Here, we discuss the NDR features in the current-voltage characteristics which originate from the electron coupling with the molecular vibrational modes. As was shown (Galperin 2005, Yeganeh 2007), the polaron formation could give rise to the NDR. This may happen if the polaron shift in the energy of the occupied state moves the energy level away from the conduction  window. As the shift depends on the electronic population, and the latter changes as the bias voltage increases, the occupied state falls out from the window between  the chemical potentials of the electrodes at certain value of the voltage. If the transport is being conducted via this occupied state, then it stops when this voltage value is reached. Correspondingly the current value drops to zero revealing a distinctive NDR feature. This is a realistic scenario based on the main features of the electron transport through molecules under polaron formation. 

In conclusion, electron-vibrion coupling could lead to the formation of a polaron which results in the energy differences between occupied and unoccupied states on the molecular bridge. At weak interaction between the bridge and the electrodes and strong electron-phonon coupling these charged and neutral states are metastable and could serve for electron conduction. This results in such interesting and potentially useful effects as the hysteresis and NDR features in the current-voltage characteristics.

\subsection{VII. Molecular junction conductance and long range electron transfer reactions}

Long range electron transfer reactions play an important part in many biological processes such as photosynthesis and enzime cathalyses (Kuznetsov 1999). Theoretical and experimental studies of these processes last more than four decades but they still remain within a very active research area. In the intramolecular electron transfer reactions the electric charge moves from one section of a molecule to another section of the same molecule. Long range transfer typically occurres in large  molecules such as proteins and/or DNA, so that these two sections are situated far apart from each other. A common setup for the transfer reactions includes a donor, a bridge and an acceptor. Due to the large distances between the donor (where an electron leaves) and the acceptor (where it arrives) typical for the long-range transfer reactions, a direct coupling between the two is negligible. Therefore, the electron participating in the transfer needs a molecular bridge providing a set of intermediate states for the electron transport.

 Essentially, the intramolecular electron transfer is a combination of nuclear environment fluctuations and electron tunneling. Electron transfer reactions result from the response of a molecule to environmental polarization fluctuations which accompany nuclear fluctuations. The molecule responds by redistribution of the electronic density thus establishing opportunities for the charge transfer to occur. The main characteristic of the electron transfer processes is the transfer  rate $ K_{et}$ which is the inversed time of the reaction. The transfer rate for the electron tunneling in the molecule interior from the donor to the acceptor depends both on the electronic transmission amplitudes and the vibrionic coupling. The latter provides the energy exchange between the electronic and nuclear systems.

Viewing  fast electronic motions in the background of the slowly moving nuclei (nonadiabatic electron transfer) and applying the Fermi golden rule of  perturbation theory, it was shown that  $ K_{et} $ could be written in the form first suggested by Marcus (Marcus 1965):
  \be
 K_{et} = \frac{2\pi}{\hbar}|H_{DA}|^2 (FC).  \label{34}
 \ee
 Here, the first cofactor is the electron transmission coefficient, $H_{DA}$ being the effective matrix element between the donor and the acceptor. The second term is the density of states weighted Franck-Condon factor which describes the effect of nuclear motions in the environment.

There exists a noticeable resemblance between the long range electron transfer and the molecular conductance, and this resemblance was analyzed in several theoretical works (Dahnovsky 2006, Mujica 1994, Nitzan 2001, Yeganeh 2007, Zimbovskaya 2002,2003). The fundamental part in both molecular conduction and electron transfer reactions is taken by the electron tunneling inside the molecule via the set of intermediate states.  
  In large molecules the electron involved in the transfer reaction or contributing to the conductance, with a high probability follows a few pathways which work as molecular bridges. The rest of the molecule do not significantly participate in the transfer/conduction and may be omitted from consideration in calculations of the electron transmission (Jones 2002, Beratan 1992). On these grounds it was suggested that the original molecule could be reduced to a simpler chain-like structure, and the latter may replace the former in calculations of molecular conductance and/or transfer rates (Daizadeh 1997). Such simplification significantly eases computations of the molecular conductance and the electron transfer rates in macromolecules.

The resemblance between the molecular conductance and long range electron transfer does not mean that these  processes are nearly identical. Along with the similarities there are substantial differences between the former and the latter. For instance, the continuum of states causing the charge transport  arises from the multitude of electronic states on the electrodes in the case of molecular conduction, and from the vibrations and fluctuations in the environment in the case of electron transfer reactions. Also, the driving force which puts electrons in motion originates from the external bias voltage applied across the junction in the case of molecular conduction. In the electron transfer situation this force appears due to the electron-vibrionic interactions in the system, or the electron transfer reaction starts as a result of photoexcitation of the donor part of the molecule. The observables, namely, the molecular conductance $ g$ and/or current $I,$ and the transfer rate $K_{et} $ differ as well.

Notwithstanding these differences the electron transport through molecular junctions and the electron transfer could be theoretically analyzed using the common formalism as recently proposed by Yeganeh, Ratner and Mujica (Yeganeh 2007). Further we follow their approach. The starting point is that one may simulate the donor-bridge-acceptor system as some kind of molecular junction, shown in the Fig. 12. Here, we represent the donor as a single state $|i\big>,$ and this state mimics the left lead (assuming the transport from the left to the right). The donor is coupled to the bridge and the latter is coupled to the continuum of the final states $|f\big> $ simulating the right lead. These couplings are described by self-energy terms $ \Delta_{i,f}.$ For simplicity we assume that there exists a single state on the bridge, and this state is coupled to a vibrionic mode of the frequency $\Omega. $ The latter  mimics  environment fluctuations. We assume that the ``leads'' are weakly coupled to the bridge $(\Delta_{L,R} \ll \lambda)$ which is typical for  electron transfer situations. Also, the representation of the environment motions by the single vibrionic mode is justified only at low temperatures when the thermal energy $ kT$ is much smaller than the electron-vibrion coupling parameter $\lambda.$

\begin{figure}[t]  
\begin{center}
\includegraphics[width=4.2cm,height=8.5cm,angle=-90]{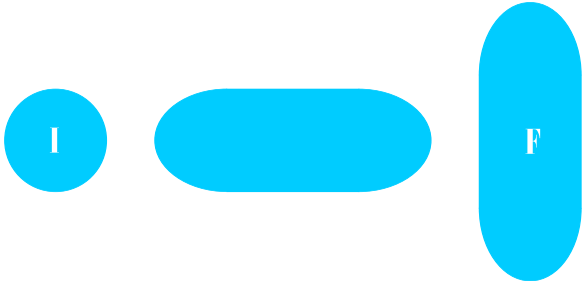}
\caption{(Color online) Schematic of the model system used in the transfer rate calculations. The initial/final reservoirs correspond to the left/right leads in the molecular junction.
}
 \label{rateI}
\end{center}\end{figure}

Now, we can write the Landauer expression for the current flowing through the ``junction'' using Eqs. (\ref{3}), (\ref{4}). Assuming that the bridge includes only one state we simplify Eq. (\ref{4}):
  \be
 T(E) = 2T_r \{\Delta_i G \Delta_f G^\dagger\} = - 2 \frac{\Delta_i\Delta_f}{\Delta_i + \Delta_f} \mbox{Im} (G).   \label{35}
  \ee
 Here, the subscripts $ i/f$ label initial and final states. Within the Fermi golden rule regime which allows to introduce the transfer rate, the bridge must be coupled to the final states much stronger than to the initial state $(\Delta_f \gg \Delta_i)$. Therefore: 
    \be
\frac{\Delta_i \Delta_f}{\Delta_i + \Delta_f}   \approx  \Delta_i  \label{36}
   \ee
 and the coupling to the final states falls out of the expressions for the electron transmission and current. Within the chosen model this is a physically reasonable result for the final states reservoir (the right ``electrode'' in our junction) was merely introduced to impose a continuum of states maintaining the transfer process at a steady state rate. Also, considering the current flow we may suppose that the initial state is always filled $[f_i(E) = 1],$ and the final states are empty $[f_f(E) = 0].$ Therefore, the current flow through the ``junction'' accepts the form:
  \be
 I = - \frac{2e}{\pi\hbar} \int dE \Delta_i \mbox{Im} (G). \label{37}
  \ee
  Both the current and the transfer rate are fluxes closely related to each other, namely: $ K_{et} = I/e. $ So, we may write:
   \be 
 K_{et} = - \frac{2}{\pi\hbar} \int dE \Delta_i \mbox{Im} (G). \label{38}
  \ee
  Now, $\Delta_i $ could be computed using the expression (\ref{7}) for the corresponding self-energy term. Keeping in mind that the ``left electrode'' includes a single state with the certain energy $\epsilon_i$ we obtain:
  \be
   \Delta_i = \mbox{Im} \left(\frac{|\tau_i|^2}{E - \epsilon_i + is}\right) =
\pi |\tau_i|^2 \delta(E - \epsilon_i).   \label{39}
  \ee
 Accordingly, the expression (\ref{38}) for the transfer rate may be reduced to the form:
  \be
 K_{et} = - \frac{2}{\hbar} |\tau_i|^2 \mbox{Im} [G(\epsilon_i)]     \label{40}
  \ee
  where $\tau_i$ represents the coupling between the donor and the molecule bridge. It must be stressed that within the chosen model, $ \tau_i $ is the only term representing the relevant state coupling which may be identified with the electronic transmission coefficient $ H_{DA}$ in the general expression   (\ref{34}) for $K_{et}.$This leaves us with the following expression for the Franck-Condon factor:
  \be 
 (FC) = - \frac{1}{\pi} \mbox{Im} [G(\epsilon_i)].  \label{41}
  \ee
 As discussed in Section IV, at weak coupling of the bridge to the leads the electron-vibrionic interaction opens the set of metastable channels for the electron transport at the energies $ E_n = \tilde E_0 + n \hbar\Omega \ (n = 0,1,2...)$ where $ \tilde E_0$ is the energy of the bridge state with the polaronic shift included. The Green's function may be approximated as a weighted sum of contributions from these channels:
   \be
 G(\epsilon_i) = \sum_{n=0}^\infty P(n) \big[\epsilon_i - \tilde E_0 - n \hbar\Omega + is \big]^{-1}    \label{42}
  \ee
    where $ s \to 0^+,$ and the coefficients $ P(n)$ are probabilities for the channels  to appear given by Eq. (\ref{18}).

Substituting Eq. (\ref{42}) into Eq. (\ref{41}) we get:
  \be
 (FC) = \sum_{n=0}^\infty P(n) \delta (\Delta F - n\hbar\Omega).   \label{43}
  \ee
 Here,  $\Delta F = \epsilon_i - \tilde E_0 \equiv \epsilon_i - E_0 + \lambda^2/(\hbar\Omega) $ is the exoergicity   of the transfer reaction, that is the free energy change originating from the nuclear displacements accompanied by the polarization fluctuations. The effect of the latter is inserted via the reorganization term $ \lambda^2/(\hbar\Omega) $ related to the polaron formation. The exoergicity in the transfer reaction takes on the part similar to that of the bias voltage in the electron transport through molecular junctions. It gives rise to the electron motion through the molecules. In the particular case when the voltage drops between the initial state (left electrode) and the molecular bridge these two quantities are directly related by $|e|V = \Delta F.$

Usually, the long range electron transfer is observed at moderately high (room) temperatures, so the low temperature approximation (\ref{43}) for the Franck-Condon factor cannot be employed. However, the expression (\ref{41}) remains valid at finite temperatures, only the expression for the Green's function must be modified to include the thermal effects. It is shown (Yeganeh 2007) that within the high-temperature limit $(kT > \hbar\Omega) $ the expression for the (FC) may be converted to the well-known form first proposed by Marcus:
  \be
 (FC) = \frac{1}{\sqrt{4\pi E_p kT}} \exp \left[-\frac{(\Delta F - E_p)^2}{4E_p kT } \right]     \label{44}
  \ee
where $ E_p = \lambda^2/(\hbar\Omega)$ is the reorganization energy. 

While studying the electron transfer reactions in practical macromolecules one keeps in mind that both donor and acceptor subsystems in the standard donor-bridge-acceptor triad are usually complex structures including multiple sites coupled to the bridge. Correspondingly, the bridge has a set of entrances and a set of exist which an electron can employ. At different values of the tunnel energy different sites of the donor and/or acceptor subsystems can give predominant contributions to the transfer. Consequently, an electron involved in the transfer arrives at the bridge and leaves from it via different entrances/exits, and it follows different pathways while on the bridge.

Also, nuclear vibrations in the environment could strongly affect the electron transmission destroying the pathways and providing  a transition to completely incoherent sequental hopping mechanism of the electron transfer. All this means that proper computation of the electron transmission factor $ H_{DA}$ for practical macromolecules is a very complicated and nontrivial task. The strong resemblance between the electron transfer reactions and the electron transport through molecules gives grounds to believe that studies of molecular conduction can provide important information concerning quantum dynamics of electron participating in the transfer reactions. One may expect that some intrinsic characteristics of the intramolecular electron transfer such as pathways of tunneling  electrons and distinctive features of donor/acceptor coupling to the bridge could be obtained in experiments on the electron transport through molecules. For instance, it was recently suggested to characterize electron pathways in molecules using the inelastic electron tunneling spectroscopy, and other advances in this area are to be expected.

\subsection{VIII. Concluding remarks}

Presently, the electron transport through molecular-scale systems is being intensively studied both theoretically and experimentally. Largely, unceasing efforts of the research community to further advance these studies are motivated by important application potentials of single molecules as active elements of various nanodevices intended to compliment current silicon based electronics. Elucidation of physics underlying electron transport through molecules is necessary in designing and operating molecular-based nanodevices. Elastic mechanisms for the electron transport through metal-molecule-metal junctions are currently understood and described fairly well. However, while moving through the molecular bridge, an electron is usually affected by the environment, and it results in the change of its energy. So, inelastic effects appear, and they may bring noticeable changes in the electron transport characteristics. Here, we concentrated on inelastic and dissipative effects originating from the molecular bridge vibrations and thermally activated   stochastic fluctuations. For simplicity and also to keep this Chapter within a reasonable length we avoided the detailed description of computational formalisms commonly used to theoretically analyze electron transport through molecular junctions. These formalisms are described elsewhere. We mostly focus on the physics of the inelastic effects in the electron transport. Therefore, we employ very simple models and techniques. 

We remark that along with nonelasticity originating from electron-vibrionic interactions which is the subject of the present review there exist inelastic effects of different kind. For instance, inelastic effects arise due to electron-photon interactions. Photoassisted transport through molecular junctions was demonstrated and theoretically addressed using several techniques. It is known that optical pumping could give rise to the charge flow in an unbiased metal-molecule-metal junction, and light emission in biased current carrying junctions could occur.  
Also, there is an issue of the molecular geometry. There are grounds to conjecture 
that in some cases the geometry of a molecule included in the junction may change as the current flows through the latter for bonds can break with enough amount of current. Obviously, this should bring a strong inelastic component to the transport, consequently affecting observables. 

This a common knowledge that electron-electron interactions may significantly influence  molecular conductance leading to Coulomb blockade and Kondo effect. To properly treat electron transport through molecular junctions one must take these interactions into consideration. Corresponding studies were carried out omitting electron-phonon interactions. However, the full treatment of the problem including both electron-electron and electron-phonon interactions is not completed so far. There exist other theoretical challenges such as the effect of bipolaron formation which originates from the effective electron-electron attraction via phonons.

Finally, practical molecular junctions are complex systems, and significant effort is necessary to bring electron transport calculations to the result which could be successfully compared with the experimental data. For this purpose one needs to compute molecular orbitals and the voltage distribution inside the junctions, to get sufficient information on the vibrionic spectrum of the molecule, the electron-phonon coupling strengths and electron-electron interactions. One needs a good quantitative computational scheme for transport calculations where all these informations are accounted for. Currently, there remain some challenges which have not been properly addressed by the theory. Therefore, comparison between theoretical and experimental results on the molecular conductance sometimes does not bring satisfactory results. However, there are firm grounds to believe that further effort of the research community will result in detailed understanding of all important aspects of the molecular conductance including inelastic and dissipative effects. Such understanding is paramount to convert molecular electronics into a viable technology.  

\subsection{Acknowledgments}

The author thanks G. M. Zimbovsky for help with the manuscript. This work was partly supported  by NSF-DMR-0934195.

\end{document}